\documentclass[natbib]{svjour3}                     % onecolumn (standard format)
\smartqed  % flush right qed marks, e.g. at end of proof
\usepackage{graphicx}
\begin{document}

\title{Locating the LCROSS Impact Craters%\thanks{Grants or other notes
%about the article that should go on the front page should be
%placed here. General acknowledgments should be placed at the end of the article.}
}
%\subtitle{Do you have a subtitle?\\ If so, write it here}

%\titlerunning{Short form of title}        % if too long for running head

\author{William Marshall \and Mark Shirley \and Zachary Moratto \and Anthony Colaprete \and Gregory Neumann \and David Smith \and Scott Hensley \and Barbara Wilson \and Martin Slade \and Brian Kennedy \and Eric Gurrola \and Leif Harcke
}

%\authorrunning{Short form of author list} % if too long for running head

\institute{William Marshall, Mark Shirley, Zachary Moratto and Anthony Colaprete \at
              NASA Ames Research Center, Moffett Field, CA 94035, USA \\
              \email{william.marshall@nasa.gov}           %  \\
%             \emph{Present address:} of F. Author  %  if needed
           \and
           Gregory Neumann and David Smith \at
              NASA Goddard Spaceflight Center, Greenbelt, MD 20769, USA
           \and 
            Scott Hensley, Barbara Wilson, Martin Slade, Brian Kennedy Eric Gurrola and Leif Harcke \at
              Jet Propulsion Laboratory, California Institute of Technology, 4800 Oak Grove Drive,
Pasadena, CA 91011, USA
}

\date{Received: date / Accepted: date}
% The correct dates will be entered by the editor

\maketitle

\begin{abstract}
The Lunar CRater Observations and Sensing Satellite (LCROSS) mission
impacted a spent Centaur rocket stage into a permanently shadowed region
near the lunar south pole. The Sheperding Spacecraft (SSC) separated $\sim$9 hours before impact and performed a small braking maneuver in order to observe the Centaur impact plume, looking for
evidence of water and other volatiles, before impacting itself. 

This paper describes the
registration of imagery of the LCROSS impact region from the mid-
and near-infrared cameras onboard the SSC, as well as from the
Goldstone radar.  We compare the Centaur impact features, positively
identified in the first two, and with a consistent feature in the third, which are interpreted as a 20 m diameter crater surrounded by a 160 m diameter ejecta region.  The images are registered to
Lunar Reconnaisance Orbiter (LRO) topographical data which allows
determination of the impact location.  This location is compared
with the impact location derived from ground-based tracking and propagation  of the
spacecraft's trajectory and with locations derived from two hybrid
imagery/trajectory methods. The four methods give a weighted average Centaur impact location of -84.6796$\,^{\circ}$, -48.7093$\,^{\circ}$, with a $1\sigma$ uncertainty of 115 m along latitude, and 44 m along longitude, just 146 m from the target impact site. Meanwhile, the trajectory-derived SSC impact location is -84.719$\,^{\circ}$, -49.61$\,^{\circ}$, with a $1\sigma$ uncertainty of 3 m along the Earth vector and 75 m orthogonal to that, 766 m from the target location and 2.803 km south-west of the Centaur impact.  

We also detail the Centaur impact angle and SSC instrument pointing errors.  Six high-level LCROSS mission requirements are shown to be met by wide margins.  We hope that these results facilitate further analyses of the LCROSS experiment data and follow-up observations of the impact region.

% \PACS{PACS code1 \and PACS code2 \and more}
% \subclass{MSC code1 \and MSC code2 \and more}
\end{abstract}

% * \section{Introduction}
\section{Introduction}

On October 9, 2009 at 11:31:19.506 UTC, a spent rocket upper stage, the Centaur,
impacted a Permanently Shadowed Region (PSR) within the crater
Cabeus near the lunar south pole.  The impact was observed by the
Lunar CRater Observations and Sensing Satellite (LCROSS) mission's
Shepherding Spacecraft (SSC) which carried nine instruments,
including cameras, spectrometers and a radiometer.  The impact was also observed from the Earth and by the Lunar Reconnaissance Orbiter (LRO) mission. The LCROSS
experiment's primary objective was to determine the presence or
absence of water ice in a permanently shadowed crater at the Moon's
South Pole and to help constrain competing water distribution
models.

% mention LRO

After launch, the LCROSS SSC controlled the Centaur for the next
four months, performing maneuvers to target the Centaur at the
planned impact site.  Approximately nine hours prior
to impact, the SSC separated from the Centaur, performed a braking
burn to build a four-minute separation between itself and the
Centaur at impact, and oriented its instruments toward the planned
impact site within Cabeus.  Fifty minutes prior to impact, the nine LCROSS
instruments were powered on and began taking data.

Initial findings based on the LCROSS data relating to the water measurements can be found in \cite{colaprete2010}, the impact
cratering process is discussed in \cite{schultz2010}, and related results from LRO are in accompanying papers. The instruments are described in more detail in
\cite{ennico2008, ennico2010, ennico2010b, heldmann2010}.  Flight operations are described in
\cite{tompkins2010a, tompkins2010b, strong2010}, the trajectory and mission design in \cite{cooley}. Overall program mission lessons learnt can be found in \cite{andrews}. Additional detail appears in the NASA Flight Director's Blog \citep{blog2009}.  

In this paper, we describe a set of interrelated post-mission analyses
that provide context for the instrument data, focusing on registering imagery of the impact region and identifying the crater and its features.  The next
section summarizes the results from these analyses, subsequent
sections then providing the full analysis.

% ** \subsection{Summary of Results}
\section{Summary of Results}

\begin{figure}[ht]
\label{impactzoom}
\begin{center}
\includegraphics*[width=0.32\textwidth,angle=0]{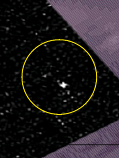}
\includegraphics*[width=0.32\textwidth,angle=0]{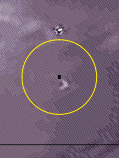}
\includegraphics*[width=0.32\textwidth,angle=0]{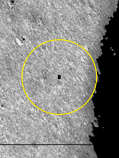}
\end{center}
\caption{Left to right: LCROSS-MIR1, LCROSS-NIR2 and Goldstone radar images of the
  Centaur impact crater. These are co-registered (see Section 3) and to the same scale. The yellow circles mark regions centred on the impact location as estimated by the trajectory information alone (see Section 4), with a radius equal to the $3\sigma$ errors of that method (423m). The black dot marks its centre.}
\end{figure}

Figure 1 shows the impact crater in Mid-Infrared
(Thermal) Camera \#1 (MIR1), Near-Infrared Camera \#2 (NIR2) and a
ground-based X-band (3 cm) radar image from Goldstone Solar System Radar (GSSR) (which has been registered to the LRO topographical data).  The relative alignments of the MIR1 and NIR2 images, taken shortly before the SSC impact, are well known due to ground calibration and registration of earlier imagery taken during the mission. The MIR1 shows a hot spot approximately $90\pm 25$ m wide. The NIR2 image shows a feature in the same location, centred approximately $-84.6774, -41.691$ degrees in Mean Earth, selenocentric coordinates. It is a a $62\pm 20$m diameter dark region, surrounded by a lighter ring that is $158\pm 40$m wide.
This is interpreted to be the crater and the ejecta blanket. Note that the estimated Centaur crater size is 25-30m based on impact modelling and constrained with LCROSS imagery \citep{schultz2010}.
The ring is brightest in the East (and slightly south).  The directional bias of the ejecta ring and the
impact angle with respect to the local terrain are both
predominantly in the easterly direction (see Section 7) suggesting one possible explanation that the impact angle caused the ejecta bias. 

A radar feature is seen in the GSSR also -- this is an area of increased radar brightness around 30\% smaller than the NIR2 feature, is thought to be due to surface roughness associated with the ejecta blanket surrounding the crater created by the impact, since it is much larger than the crater itself. Table \ref{summarytable} summarizes the averaged location and basic features for the Centaur and SSC impact sites and other results from this paper.  Subsequent sections detail the derivation of each quantity.  

\begin{table}[h]
\begin{tabular}{ p{4cm} p{3cm} p{3cm} }
\hline
Feature Characteristic&Centaur Impact&SSC Impact\\
\hline
\hspace{0 pc}Time (UTC)&11:31:19.506&11:35:36.116\\
\hspace{0 pc}Lattitude ($\,^{\circ}$)& $-84.6796 \pm 0.004$&$-84.729\pm 0.001$\\
\hspace{0 pc}Longitude ($\,^{\circ}$)&$-48.7093 \pm 0.016$&$-49.61\pm 0.014$\\
\hspace{0 pc}Altitude (m)&$-3,825 \pm 10$&$-3,809 \pm 10$\\
\hspace{0 pc}Error in Lat. $\times$ Long. (m)&$119 \times 46$&$31 \times 39$\\
\hspace{0 pc}Dark Region Diameter (m)&$62\pm 20$&-\\
\hspace{0 pc}Light Ring Diameter (m)&$158\pm 40$&-\\
\hspace{0 pc}Impact Angle ($\,^{\circ}$)&$3.67 \pm 2.3$&-\\
\hspace{0 pc}Distance to Target (m)&146&766\\
\hline
\end{tabular}
\caption{Key results of locations estimates and impact craters for the SSC and Centaur. Errors given are $1\sigma$. Altitude is with respect to a mean lunar radius
  of 1737.4 km, and latitude and longitude are given in Mean Earth, selenocentric coordinates times are in UCT on 9th October 2009.}
\label{summarytable}
\end{table}

%SSC Impact diameter dark region19\pm 10$

%NOTE: removed this from the caption - The NIR image was taken at
%11.35.31.401

%Subsequent sections cover these topics:
%\begin{itemize}
%\item Section \ref{sec:registration}: Registration of Key Images
%\item Section \ref{sub:centaurfeature}: Centaur Impact Feature
%\item Section \ref{sub:sscfeature}: SSC Impact Feature
%\item Section \ref{sec:centaurlocation}: Centaur Impact Location
%\item Section \ref{sec:ssclocation}: SSC Impact Location
%\item Section \ref{sec:impactangle}: Impact Angle
%\item Section \ref{sec:timinganalysis}: Timing Analysis
%\item Section \ref{sec:requirements}: Mission Requirements and Lessons Learned
%\item Section \ref{sec:conclusion}: Conclusions
%\end{itemize}

% * \section{Registration of Key Images}
\section{Registration of Key Images and Impact Feature Identification}
\label{sec:registration}

This section describes the process of registering images of the
Centaur impact region against each other and the identification within the imagery of the impact feature.  Passive imaging of the
impact site is difficult because it lies in permanent Sun-shadow.  However, the MIR cameras could see the impact location due to the thermal signature of impact heated regolith. In addition, sufficient detail for registration is visible in some NIR2 camera images shortly
before the SSC's impact because at this point the camera's field of view was
contained fully within the PSR so the dynamic range of each image is
reduced and its signal-to-noise ratio is improved.

Naturally, active imaging methods do not suffer from the same issue and thus LIDAR images
from LRO (from the mini RF instrument) and GSSR images are available. The latter is possible because although in permanent Sun shadow, the floor of Cabeus crater is occasionally in Earth view. In total four
data sets are used in the registration here: imagery from LCROSS's Mid-Infrared Camera 1 (MIR1)
and Near-Infrared Camera 2 (NIR2) shortly before SSC impact; Goldstone
radar before (May 2009) and after (November 2009) impact observations; and LRO's Lunar Orbiting Laser Altimeter (LOLA)
topographical data. (Note that LRO mini-RF data were also registered but due to its lower spatial resolution compared to GSSR, it is not shown in the present analysis.)

The registration process follows five steps. The first step is to identify the hotspot in the MIR1 image. This is relatively simple because the average temperature of the surface within the PSR is
considerably below the camera's detection threshold, residual heat
from the Centaur impact is the only visible feature that moves from image to
image -- as the spacecraft frame moves over the lunar surface -- over the last 30 seconds before communication was lost with
the SSC.  All other features remain fixed, e.g., hot pixels and
faint ghosting from earlier images that persists until the next,
automated flat field correction within the camera.

%NOTE: removed this from the caption - The NIR image was taken at
%11.35.31.401

% The impact feature appears in one NIR2 image.  It would be just
% out of frame in the next image.  It's not clearly visible in the
% previous NIR2 image, however many features come and go.  Our
% interpretation of this is that the tail end of the ejecta cloud is
% still falling back to the surface, and portions of that cloud are
% within the field of view and obscure some surface features.

% The crater floor is periodically in Earth view and thus radar
% imaging can be performed from the ground.

% Point not made: the crater floor was not in view from earth the
% morning of the impact.

% Covered in the image caption:
% The impact site of the Centaur and SSC based upon the trajectory
% based impact location estimates, are marked on the LOLA
% topographical map (Figure \ref{nir-lola}).  The marks are circled
% by $3\sigma$ trajectory errors.

The second step is to identify the Centaur crater feature in the NIR2
image. The relative pointing of the NIR2 and MIR1 cameras are known from pre-launch calibration and, in addition, can be verified and updated by registration of overlapping images post launch (the latter method was used as the basis of alignment herein). However, because the LCROSS camera exposures were not time synchronized, we take the MIR1 images immediately before and after each NIR2
image, identify the hotspots in each, and interpolate between them
to create a hotspot location corresponding to the time of the NIR2
image.  In the case of the NIR2 image shown in Figure 1 (timestamp 3460653481, corresponding to 11:35:31.401 on 09 Oct, 2010, approximately 4s prior to it's own impact), the interpolated hot spot in the MIR1 image lies directly on top of the impact feature seen in the NIR2 image, as is clear in the Figure.

%From left to right: before impact image; after impact image and difference image. The yellow circles represent the $3\sigma$ error estimates around location estimate of the Centaur (upper right) and SSC (lower left) impact locations based on trajectory only. The impact feature seen just below the center of the upper circle is at a location that is consistent with the trajectory-derived impact location with errors, but is not statistically significant in the difference image.

\begin{figure}[t]
\label{nir-lola}
\begin{center}
\includegraphics*[width=0.49\textwidth,angle=0]{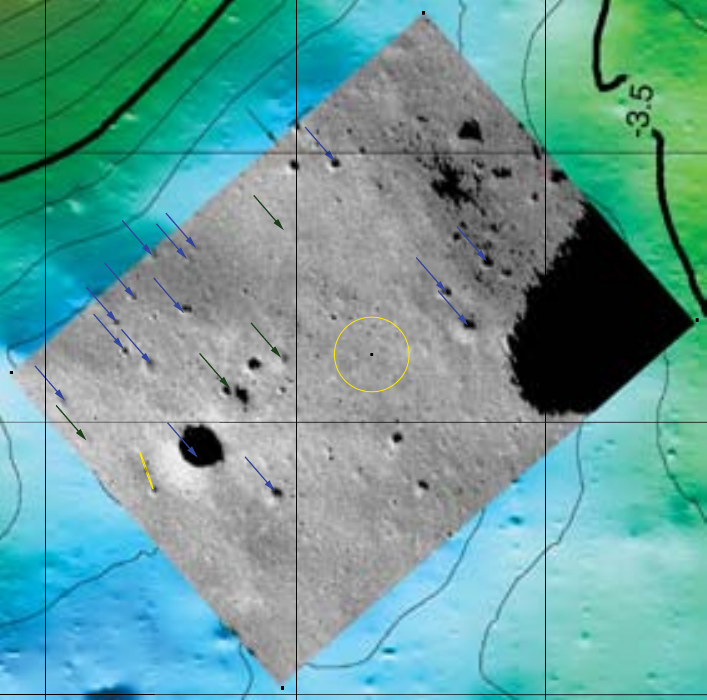}
\includegraphics*[width=0.49\textwidth,angle=0]{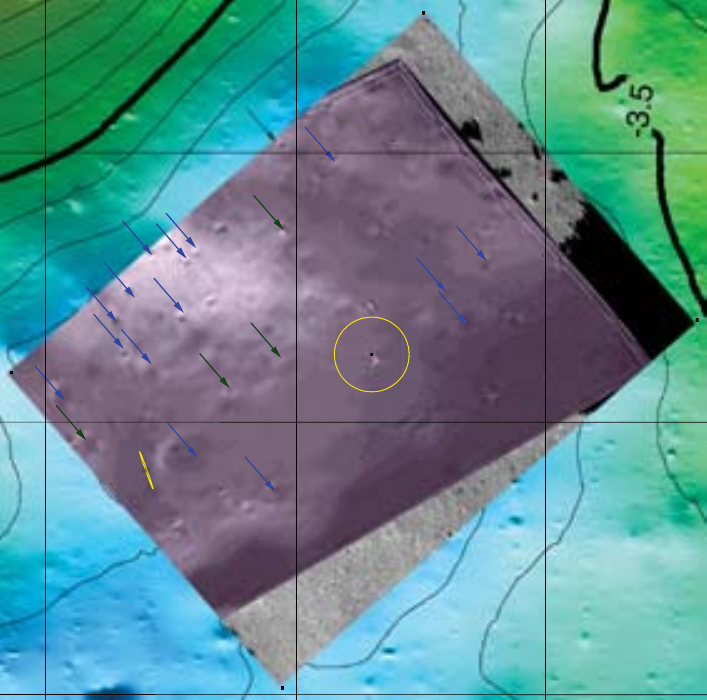}
\end{center}
\caption{
GSSR X-band (3 cm) radar imagery overlaid on LRO-LOLA map (left) and with both the GSSR and LCROSS-NIR2 image
  overlaid (right). The blue arrows represent markers used for
  registration. The green arrows are additional features found to
  line up upon registration. The yellow circle represents the
  $3\sigma$ error around the location estimate of the Centaur
  impact (upper right) and SSC impact (lower left), each based on trajectory data from tracking (Section 4). These errors correspond to a radius of 423m in both latitude and longitude for the Centaur and 9x225m in the Earth vector (projected onto the lunar surface at the impact location) and orthogonal to the Earth vector for the SSC. The grid lines are at 1$\,^{\circ}$ intervals in longitude ($\sim 3.03$ km) and 0.1$\,^{\circ}$ intervals in latitude ($\sim 2.82$ km).}
\end{figure}

The third step is to register the NIR2 image to the GSSR radar data (in particular we use the exposure taken before the Centaur impact). To do this, the NIR2 image was approximately
aligned with the GSSR image based on the
spacecraft position and attitude that the time of the image
was taken, together with camera pointing on the spacecraft, providing approximate latitude and longitude locations of the corner pixels. Then, eight features were identified that were present in
both the NIR2 and GSSR images.  Markers were placed in the centre of these features,
and an elastic warping process (size, orientation, stretch and deformation) was then used to shape the NIR2 image onto
the GSSR image.  This was done with the BunwarpJ plug in for ImageJ
\citep{Arganda-Carreras}.  After the images were registered, more
than seven additional features were found to be common to both images,
giving confidence in the resulting registration. This was repeated now including all 15 common features as markers allowing a more precise registration. Again a further 5 features were found to be in common. Figure 2 shows the registered images and aligned features.

The fourth step is to place the combined registered image set (MIR1, NIR2 and GSSR imagery) onto the LOLA terrain.  This is straightforward since the GSSR data has already been registered to the LOLA data, and corner pixel coordinates were used to align the image. 

% ** \subsection{Centaur Impact Feature}
%\subsection{Centaur Impact Feature}
%\label{sub:centaurfeature}

The fifth step is to identify the impact feature in the data sets.
To do this, the trajectory-based impact location (see Section 4) was marked onto the registered set of images. An ellipse was drawn to mark the 3-sigma errors associated with the trajectory estimate. The MIR1 image clearly shows a hotspot near the centre of the trajectory ellipse (see Figure 1). Moreover this is the aforementioned one feature that moves from image frame to frame, thus confirming it is the heat signature of the Centaur impact rather than noise. 

\begin{figure}[t]
\label{diff}
\begin{center}
\includegraphics*[width=0.99\textwidth,angle=0]{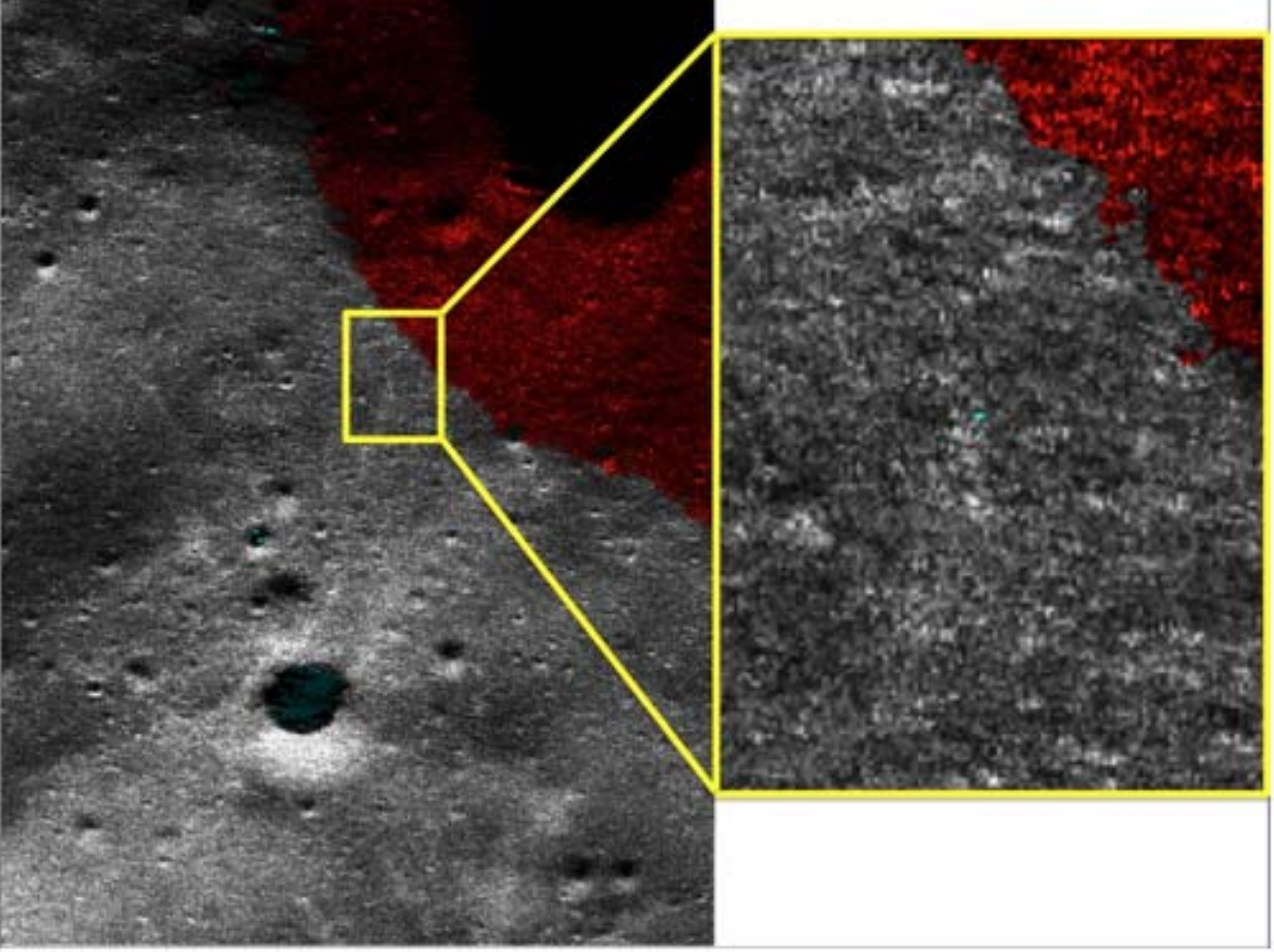}
\end{center}
\caption{
Goldstone X-band radar image of the Centaur impact region taken post impact and backprojection processed as described in the text. The blue spot shows the trajectory impact location as determined by a CFAR detection algorithm described in detail in the text. 
The red pixels represent areas which the CFAR program detected to be above its threshold in 
the ratio change image (constructed from before and after images). The red pixels are predominantly associated with the di?erence in shadows due to the di?erent viewing angles for the 
before and after impact exposures, however there is a small feature approximately 75m from 
the trajectory-based impact location. This feature has a $83\%$ conÞdence level and has a location 
consistent with the Centaur impact.}
\end{figure}

Precisely aligned with this in the NIR2 image is a feature described in Section 2. For the detection in the GSSR data, the process was slightly more involved. The data were processed through a backprojection image formation processor to 5 m range resolution and 10m cross-range or azimuth resolution.  The data were orthorectified during image formation by backprojecting onto the digital elevation model of the moon from the Japan Space Agency LALT laser altimeter instrument on the Kaguya lunar mission.  The data were formed on an orthographic projection grid (note: this is not polar stereographic).  A region 15 km on a side was extracted from the pre-impact (May 2009) and post-impact (November 2009) data, approximately centred on the LCROSS impact site.  These single-look complex (SLC) data were averaged incoherently in range by a factor of 2 to form square (10 m by 10 m) pixels.  These imagery were further de-speckled by convolving a unity gain boxcar filter 40m on a side (or 4 by 4 pixels at 10 m resolution) with the data.  The resulting imagery, now at approximately 40 m resolution (but still at 10 m posting) were input to a constant false alarm rate (CFAR) thresholding program.  A ratio change image was formed by dividing the November data by the May data, pixel by pixel.  The resulting ratio image was fit to a log-normal probability distribution.  The region chosen for this fit was the lower half of the image, or a 7.5 by 15 km region, which did not contain the putative impact zone.  Once the log-normal parameters were extracted, an operating threshold of 10 false alarms per square kilometer was chosen, and the ratio image thresholded to highlight those pixels which contained statistically significant differences in reflectivity.  This CFAR detector highlighted a point ~75 m away from the impact point indicated by radio tracking of the final trajectory of the LCROSS impactor.  From 2-D probabilistic analysis, for 10 FA/km2, the probability that a random event lands within 75m of the impact zone is $17\%$.  Restating this result, there is $83\%$ confidence in the result, or approximately a 1 out of 5 chance that the detected point was a random event. The processed post-impact GSSR data around the LCROSS impact region is shown in Figure 3 (note: this image is different from the GSSR data shown in the other figures since it had different processing, as described above). This cannot be taken as a statistically significant confirmation that this is the Centaur crater but it is consistent with it. 

Further analysis of the GSSR data could lead to a more deÞnitive assessment of this 
feature. Indeed, to date only one exposure before and one after have been processed, whereas six before and four after exposures were taken. Combining these could improve the signal-to-noise ratio (particularly the radar speckle component) and allow clearer identification of the change feature. Further, of the available imagery, only the radar can potentially contain an SSC crater feature and thus this may enable verification of the SSC location.

\begin{figure}[t]
\label{nir-radar-lola}
\begin{center}
\includegraphics*[width=0.99\textwidth,angle=0]{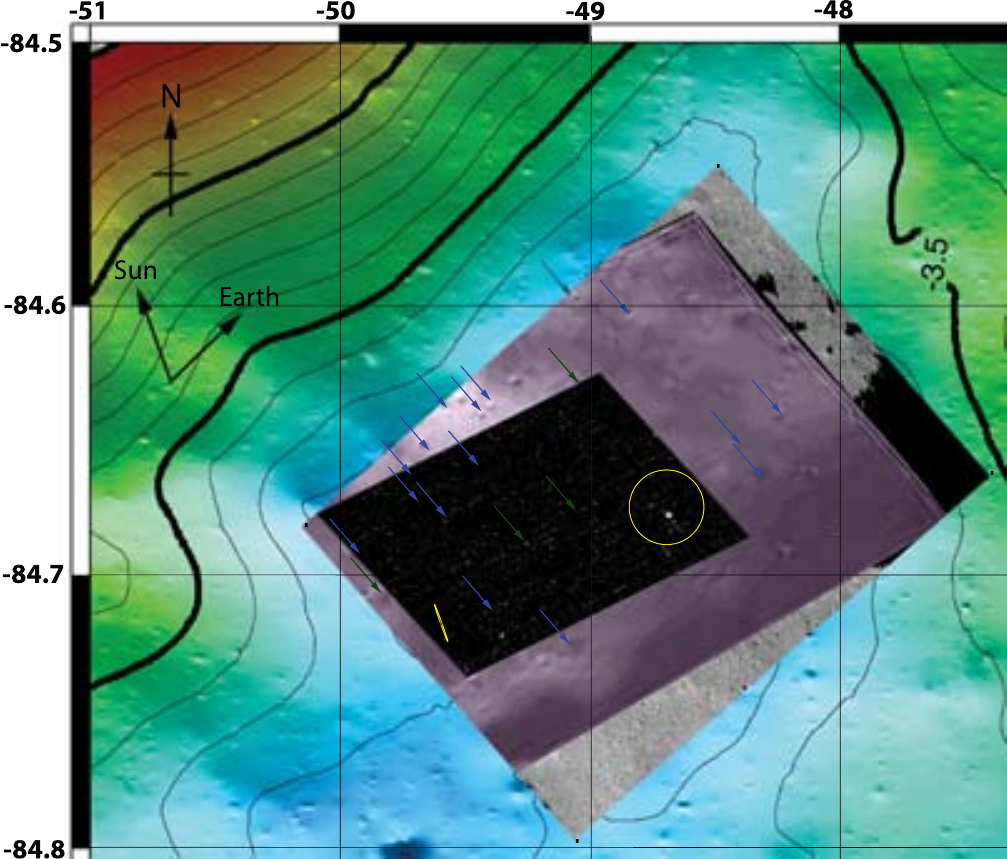}
\end{center}
\caption{LCROSS-MIR1, LCROSS-NIR2, Goldstone Radar and LRO-LOLA images of the
  impact site co-registered. The grid lines are at 1$\,^{\circ}$ intervals in longitude ($\sim 3.03$ km) and 0.1$\,^{\circ}$ intervals in latitude ($\sim 2.82$ km). The sun and Earth vectors at the time of the Centaur impact are shown (note that these vectors are different for the radar image which was taken about a month after the impact. The yellow circles and the arrows are the same as for Figure 2.}
\end{figure}

The complete set of MIR1, NIR2 and GSSR images registered to the LOLA background is shown in Figure 4. Four factors increase our confidence in the image alignment and the
location of the Centaur impact feature: (1) the alignment of $\sim20$ features
between any two registered image pair (NIR2-GSSR, NIR2-LOLA and GSSR-LOLA); (2) the good alignment of the Centaur impact feature shown in the NIR2 and the MIR1 hot spot; and (3) the radar feature in the $3\sigma$ error margins of the trajectory-derived impact location of the Centaur (albeit not statistically significant). The agreement of this combination of imaging types increases our confidence in the identification of the impact site.

% ** \subsection{SSC Impact Feature}
%\subsection{SSC Impact Feature}
%\label{sub:sscfeature}

% * \section{Centaur Impact Location}
\section{Centaur Impact Location}
\label{sec:centaurlocation}

\begin{table}[t]
\caption{Methods for locating the Centaur impact site}
\begin{tabular}{ p {2.5cm} p{8.5cm} }
\hline
Name&Description\\
\hline
Trajectory Only&Based solely on orbit determination from the Deep Space Network (DSN) tracking
and propagation in the lunar gravitational field\\
Imagery Only&Registration of NIR2 and GSSR imagery that has the impact
Crater in view against LOLA-derived reference imagery\\
Hotspot Forward Projection&Hybrid approach (manual): hotspot forward projection to the
lunar surface of the impact location hotspot in MIR1 as determined by SSC position,
attitude and camera pointing data\\
Bundle-Adjustment&Hybrid approach (automated): bundle-adjusted (BA) image
registration against a LOLA-derived reference prior to impact,
utilizing the trajectory and spacecraft attitude data\\
\hline
\end{tabular}
\label{method-table}
\end{table}

The Centaur impact location was estimated using the four,
partially-independent techniques summarised in Table
\ref{method-table}.  Here we introduce all four methods and their
results.  A complete list of the data used and an overview of the information flow for each of these methods is given in Appendix A. Each method of impact location determination is described in full
detail in Appendix B. In summary, the image only approach comes directly from the impact registration of Section \ref{sec:registration} that provides a location of features seen in the images on the lunar surface. The trajectory only method uses just the DSN tracking data (Doppler and range data) along with propagation of that trajectory in the lunar gravity field, and other necessary perturbing forces to predict an impact location on the lunar surface, and no imagery at all. The hybrid approaches use the same imagery but not the registration itself, together with the trajectory data in order to estimate the impact location. Appendix B includes full details of the trajectory states (and propagation techniques) used that may be of use to other research efforts, whilst further information on the trajectory design and optimisation techniques use in LCROSS can be found in \cite{cooley}.

% NOTE: Why are we deferring this info flow?  The main thing I
% wanted was the summarize the results early.  We've done that.  By
% breaking this up further, we just return to the subject one more
% time.

\begin{table}[b]
\caption{Summary of the Centaur impact location, using four
  different methods. Altitude is with respect to a mean lunar radius
  of 1737.4 km, and latitude and longitude are given in Mean
  Earth, selenocentric coordinates. The trajectory estimate given is that which includes the outgassing adjustment (see Appendix B)}
\begin{tabular}{ c c p{2.5cm} p{2cm} p{1.5cm} c}
\hline
Method&Lat ($\,^{\circ}$) &Long ($\,^{\circ}$) &Alt. (km) &Error (m)\\
\hline
1. Image Only&-84.6774&-48.691&n/a&130x130\\
2. Trajectory&-84.6827&-48.6688 &-3.829&$141\times 141$\\
3. Hybrid: Manual  & -84.6805&  -48.7399& n/a &$45\times 87$\\
4. Hybrid: Auto.  &-84.6782  &-48.7214 &-3.816 &$66 \times 102$\\
\hline
Weighted Ave. &$-84.6796 \pm 0.004$&$-48.7093 \pm 0.016$&$-3.825 \pm 0.01$&$115 \times 44$\\
Target&-84.675&-48.725&-3.827	&\\
\hline
\end{tabular}
\label{method-results-table}
\end{table}

Table \ref{method-results-table} summarizes the results from each of
the four methods, including the $1\sigma$ errors. The final weighted average location is estimated to be -84.6796 degrees north, -48.7093 degrees west, in Mean Earth, selenocentric coordinates. This equates to 106.074 km, -120.85 km, -1726.1 km in Cartesian coordinates. Figure
5 shows these results for the Centaur
on the right and, separately, for the SSC on the left.  (The process
for the SSC is described in Section 4).  

The distance from the weighted average location to the target location is 139 m North and 44 m East, 146 m in total.  The $1\sigma$ error on this is 115m along latitude and 44m along longitude.  Thus the worst case mean plus $3\sigma$ range is $< 500$ m.

% ** \subsection{Centaur Trajectory Only}

% ** \subsection{Combined Centaur Location Error Analysis}
\subsection{Centaur Location Error Analysis}

The four methods were combined by a weighted average to generate the
impact location in Table \ref{method-results-table} above.  The
relative weights are inversely proportional to the error of each
measurement.  The errors given are $1\sigma$ assuming a normal
distribution of measurements, and equate to a
distance of 115 m from the mean position in the North-South direction and 
44m in East-West.  An
alternative statement of the errors is that all estimates lie within
a 148m radius circle centred on the average.

\begin{figure}[]
\label{impactsummary}
\begin{center}
\includegraphics*[width=5 cm,angle=0]{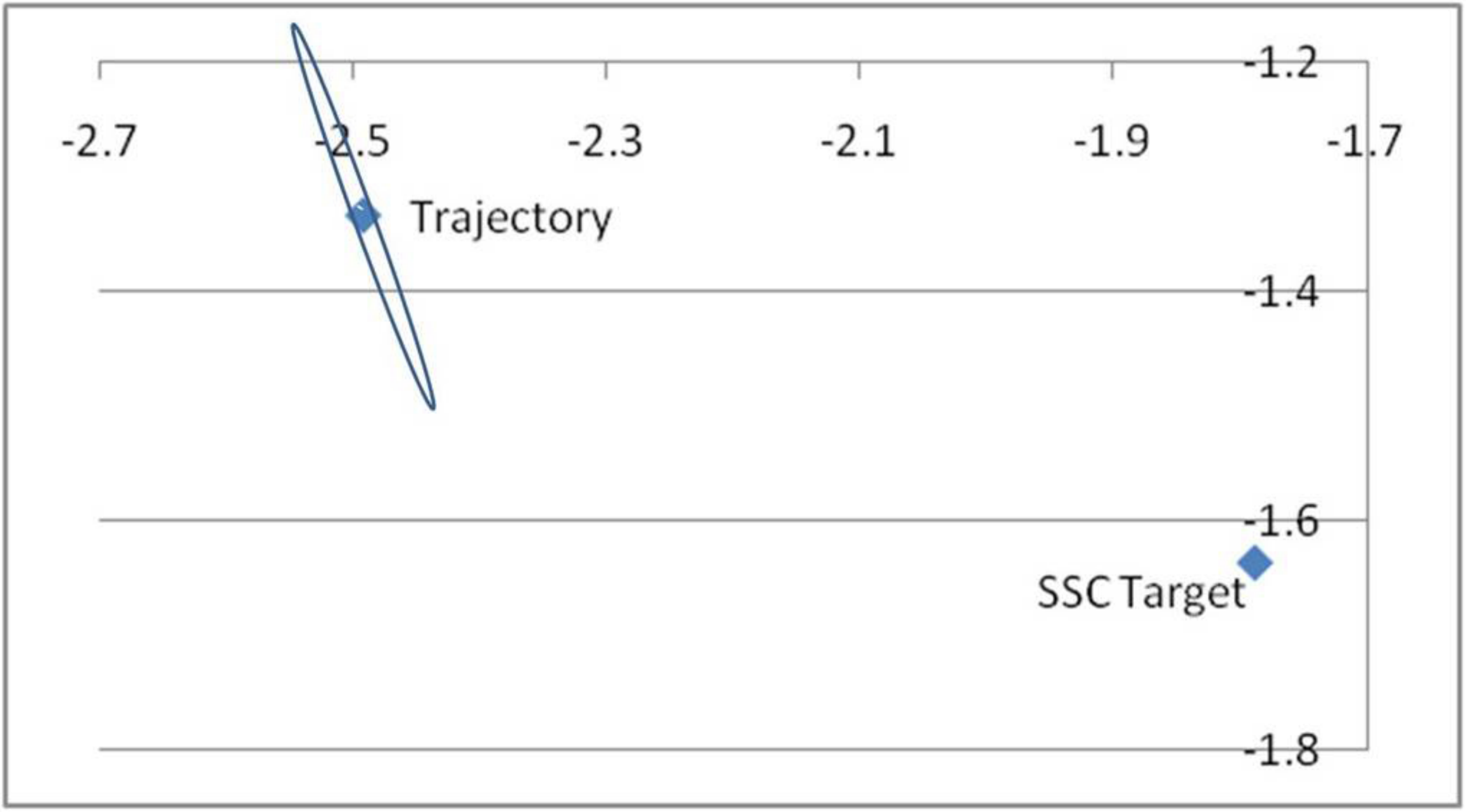}
\includegraphics*[width=7 cm,angle=0]{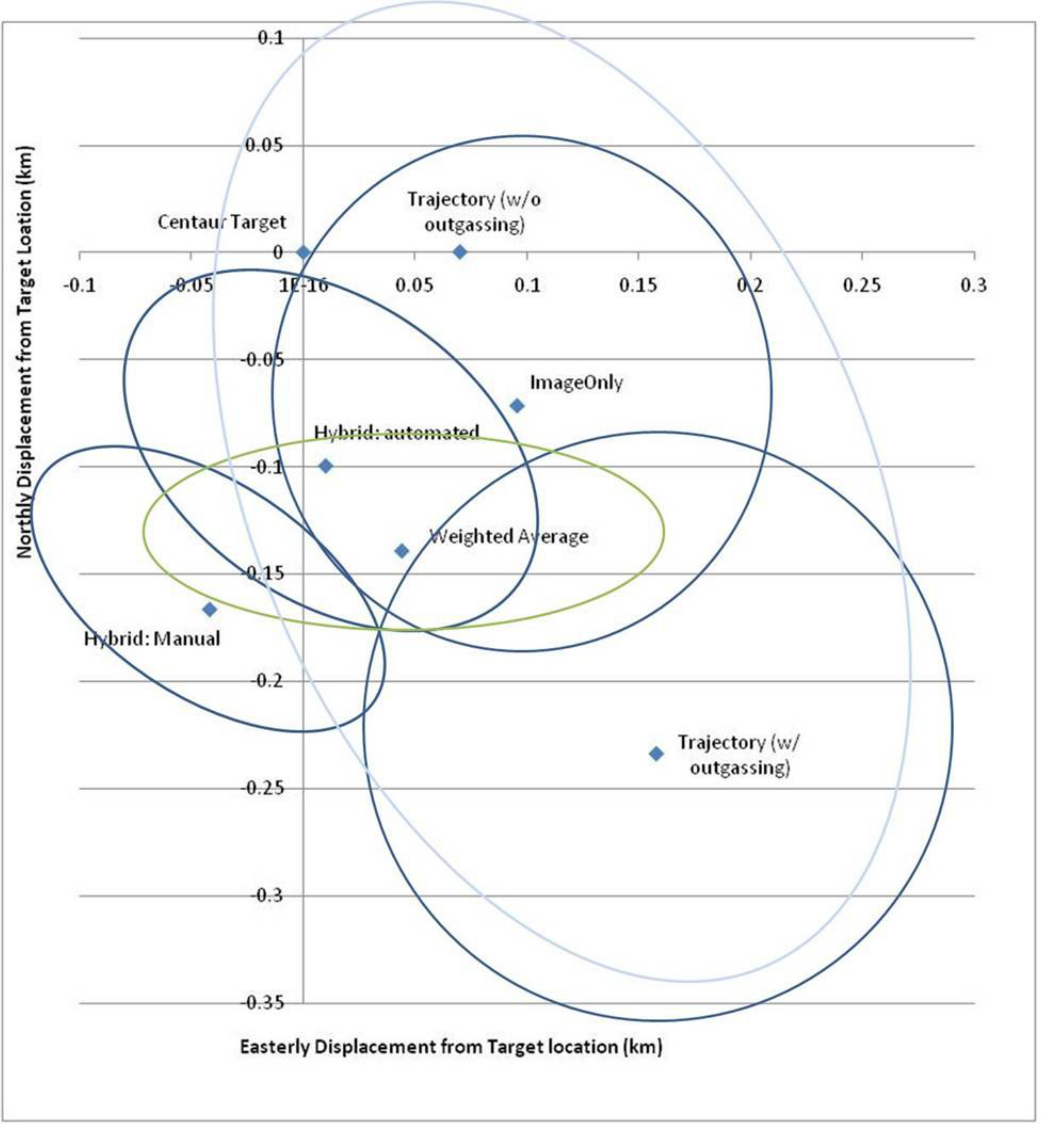}
\end{center}
\caption{SSC and Centaur impact locations estimates compared. The two figures show the
  clusters of estimates for the SSC (left) and Centaur (right) using
  their 2 and 4 methods respectively, as well as the original
  target coordinates. All distances are in kilometers.  The origin of
 both images is the Centaur impact target location.  The ellipses
  represent $1\sigma$ errors for each method - each centred on
  their respective method, apart from the Centaur trajectory method
  without outgassing whose error ellipse (pale blue) has a bias
  towards the south-south-east (the anti-sunward direction). The right hand image also shows the overall error ellipse in green centred on the weighted average impact location.}
\end{figure}

Figure 5 makes clear that all the location methods are just about mutually consistent at the $1\sigma$ level. A tightening of the error ellipsoids may be possible from a variety of sources: (1)
imagery of the crater from LRO/LROC, LOLA or MiniRF; (2) improved imagery from GSSR allowing confirmation of the Centaur (and even SSC) impact features in that data; (3) image bundle adjustment registration which also refines the spacecraft trajectory and attitude and thus
updates the trajectory based approaches; or (4) updated timing
information on the spacecraft electronics boards.

%  or ground based radar providing independent coordinate determination

% * \section{SSC Impact Location}
\section{SSC Impact Location}
\label{sec:ssclocation}

The SSC impact location was only possible to estimate using the trajectory method. The final weighted average SSC location is estimated to be -84.719 degrees north, -49.61 degrees west, in Mean Earth, selenocentric coordinates. This equates to 103.364 km, -121.54 km, -1726.21 km in Cartesian coordinates. The results are
summarised in Table \ref{ssc-method-results-table}. The analysis is described in full
detail in Appendix C.

\begin{table}[t]
\caption{Summary of the SSC impact location, using two different
methods. Latitude and longitude are given in Mean Earth, selenocentric coordinates}
\begin{tabular}{ p{2.2cm} c p{2.5cm} p{1.5cm} p{1.4cm} c}
\hline
Method&Lat ($\,^{\circ}$) &Long ($\,^{\circ}$) &Alt. (km) &Error (m)\\
\hline
Trajectory&-84.719&-49.61&n/a&$3\times 75$\\
\hline
Target&-84.729&-49.36&-3.8091&\\
\hline
\end{tabular}
\label{ssc-method-results-table}
\end{table}

The SSC was tracked until 1 hour before its impact, whereas the
Centaur was untracked after separating from the SSC, thus the SSC has more precise trajectory data.  However, the Centaur crater was imaged by the SSC
cameras and as a consequence additional methods could be
used to locate the Centaur impact, while only one was used in to
locate the SSC impact (see Appendix A). Also, the Centaur crater
produced a larger crater, easier to see in the radar image.

The distance from the trajectory location to the target location is $\sim$766 m.  The errors on the SSC trajectory as calculated by the JPL Orbital Determination team are 3 m x 75 m, $1\sigma$,
where the 3m is in the Earth vector (as projected onto the lunar
surface) and the 75 m is orthogonal to that. These are dominated by
the errors of the tracking process. Thus the worst case mean plus $3\sigma$ range from target impact location is $\sim1000$ m. This is 2.803 km south-west of the Centaur impact.

% ** \subsection{SSC Impact Error Analysis}
\subsection{SSC Impact Error Analysis}
\label{sub:SSCErrorAnalysis}

The averaged SSC location is almost 800 m from the target
SSC location.  This is compared to 5x smaller distance for the Centaur estimated impact point from its target.  This can be explained
by the fact that during the final mission events, and in particular
the separation event, utmost priority was given to placing the
Centaur close to its target, and sometimes this was at the expense
of the SSC target accuracy. The separation event $\Delta V$
was used as a final maneuver to correct the Centaur trajectory in
order to hit the target.  The separation event was originally
planned to be along the velocity vector but was taken off that in
order to do a small correction thus causing a shift in the SSC trajectory. 

The largest source of error in the SSC impact location should have been due to the 9 m/s braking burn that took place 40 minutes after separation.  Other error sources include the large amount of attitude control thrusting during the various slews, as well as the tight dead band and more frequent quaternion updates used during observations.  We estimated the braking burn error to be about -1$\%$ based on the post-burn OD (i.e., a slightly cold burn). The trajectory and image only analysis methods have small errors associated with their location estimates relative to the distance from the target and are in close agreement. The errors on the SSC trajectory as calculated by the JPL Orbital Determination team are 3 m x 75 m, $1\sigma$,
where the 3m is in the Earth vector (as projected onto the lunar
surface) and the 75 m is orthogonal to that. These are dominated by
the errors of the tracking process. 

% * \section{Pointing Accuracy}
\section{Pointing Accuracy}
\label{sec:PointingAccuracy}

The Near Infrared Spectrometer 1 (NSP1) was the primary water-detection
instrument on LCROSS.  Knowing the Centaur impact location allows
the accuracy with which the SSC pointed this spectrometer at the
impact to be evaluated.  Figure 6 shows the angle
between the NSP1 boresight and the Centaur impact location over the
last 15 minutes of the mission.  The SSC kept the Centaur impact
location continuously within the 0.1$\,^{\circ}$ of the NSP1
field-of-view (FOV) from 5 minutes prior to Centaur impact until 3
minutes after. The angle remains under 0.5$\,^{\circ}$ until 11:35:07
UTC, (27 seconds before loss-of-signal).  The SSC
kept the Centaur impact location within the MIR1 Camera
fields-of-view until 11:35:32, (2 seconds before loss-of-signal). The maximum error before the SSC starts to deviate from the Centaur
impact location is 0.198$\,^{\circ}$.  

\begin{figure}[t]
\label{nsp-angle}
\begin{center}
\includegraphics*[width=0.75\textwidth,angle=0]{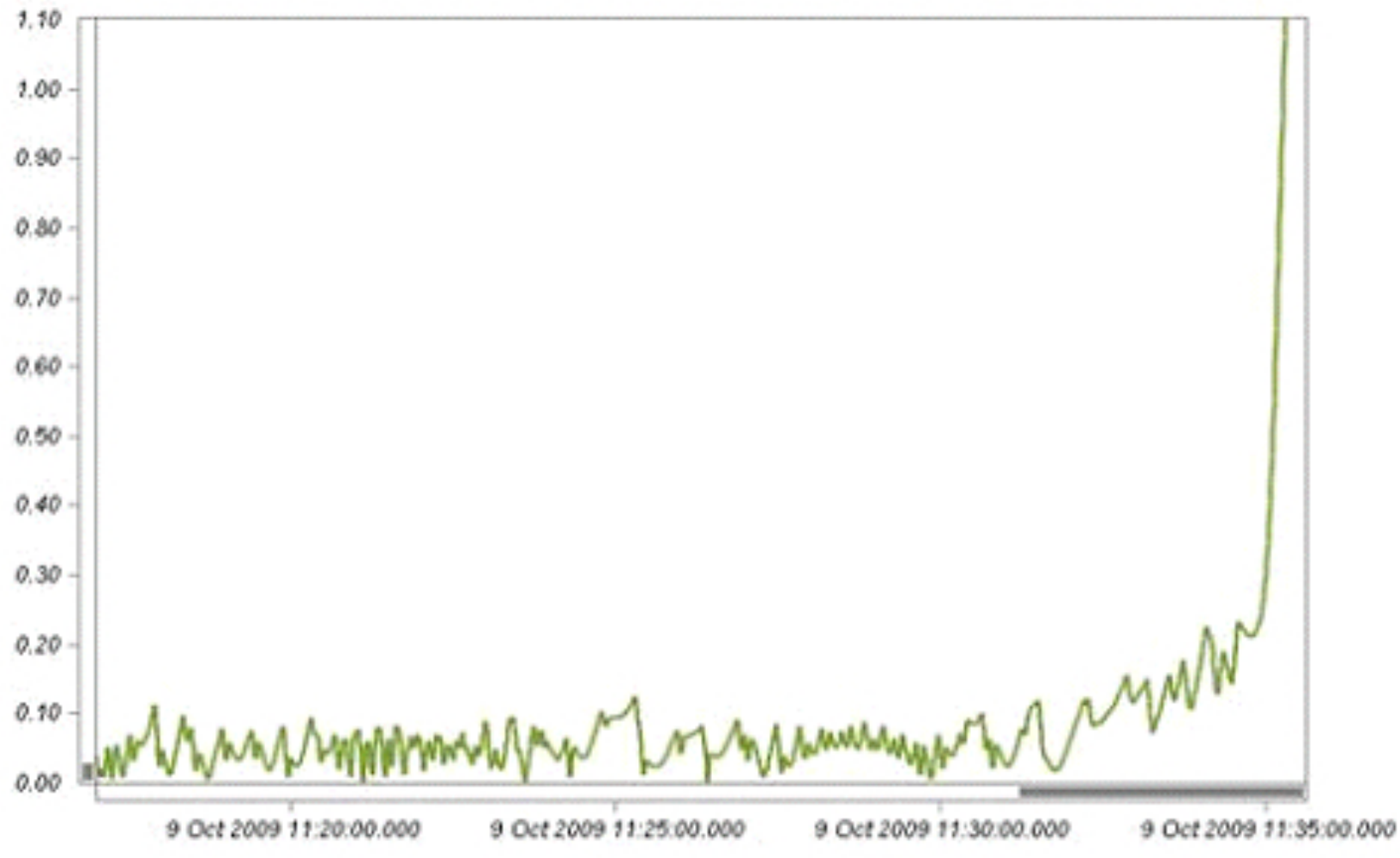}
\end{center}
\caption{NSP1 boresight angle from Centaur impact position in degrees as a function of time}
\end{figure}

\
% * \section{Impact Angle}
\section{Impact Angle}
\label{sec:impactangle}

\begin{figure}[b]
\label{slope}
\begin{center}
\includegraphics*[width=0.75\textwidth,angle=0]{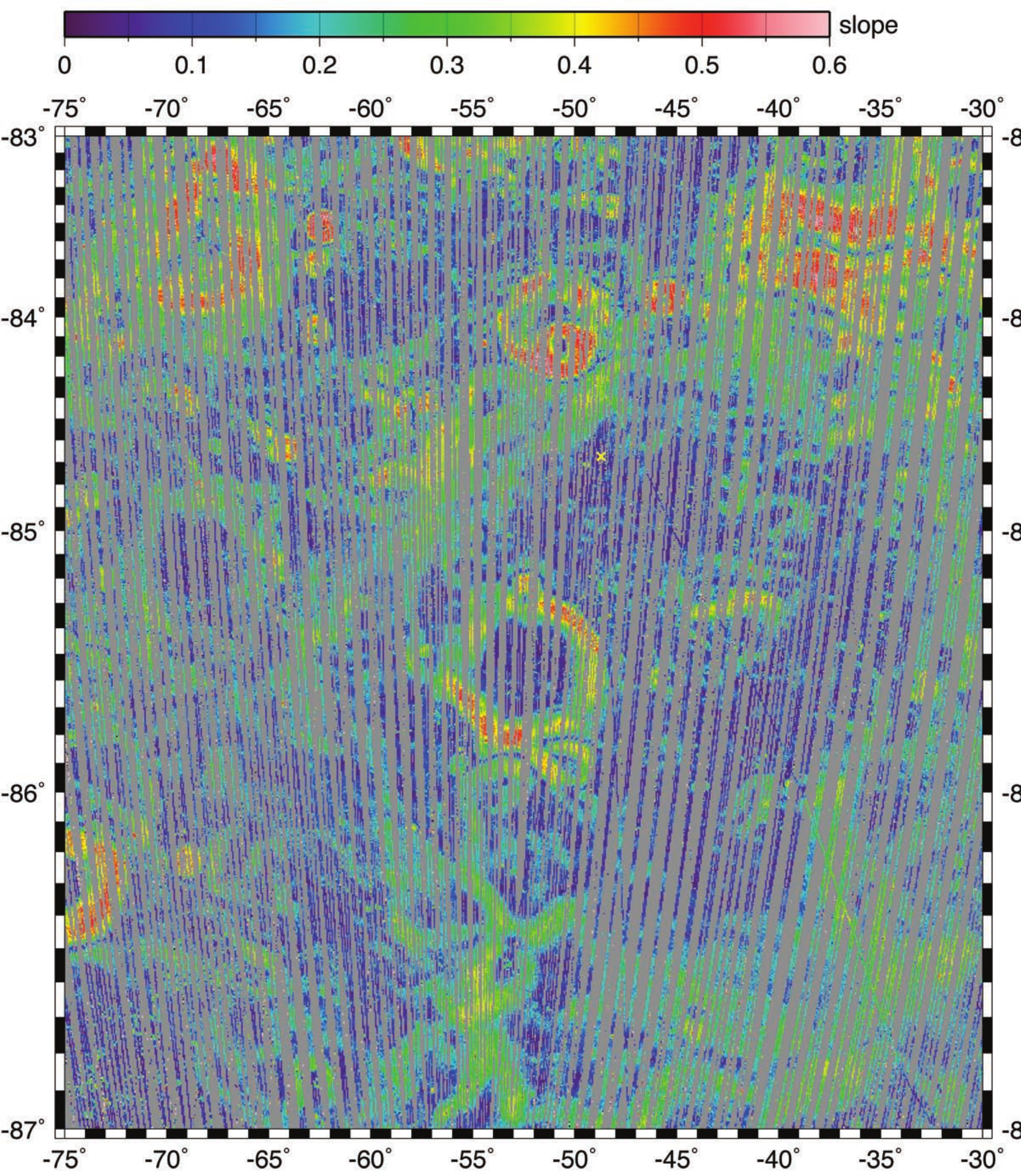}
\end{center}
\caption{LOLA Slope Map. The yellow cross indicates the Centaur
  Impact Location.  The scale is linear with a slope of 0 translating to 0 $\,^{\circ}$ and a slope of 0.6 to 30.96
  $\,^{\circ}$.}
\end{figure}

There are two components to finding the impact angle of the Centaur
with respect to the local terrain slope: (1) the velocity vector of the
Centaur at impact, as measured with respect to the surface of a perfect
lunar sphere; and (2) the slope at the impact point given the
local terrain.

The most accurate trajectory model that we have post impact was used
to generate the local Centaur velocity vector at impact, using
STK. At impact, the local lateral velocity components are -0.00217$\,^{\circ}$/s (= -0.06574 km/s) in latitude and 0.058947$\,^{\circ}$/s (=0.165944 km/s) in longitude. Given the total
velocity of 2.506885 km/s (or local altitude rate of 2.50052 km/s)
at impact, this implies a latitude impact angle of -1.50271$\,^{\circ}$
and longitude impact angle of 3.79549$\,^{\circ}$. This is obtained by projecting the trajectory
  onto a plane of constant longitude and measuring the angle between
  normal and trajectory in that plane. A negative angle is defined as
  the Centaur velocity being in the negative latitude direction. The total impact
angle is thus 4.08295$\,^{\circ}$ from the normal to the lunar
surface (or, 85.917$\,^{\circ}$ from horizontal) if the local slope
was zero. Thus the Centaur was heading in a South-Easterly (mainly Easterly)
direction at impact.

LOLA topographical data were used to generate a local slope at the
impact site by measuring the altitude at the impact site as well as
at adjacent nearest neighbour points in the north-south axis and the
east-west axis. The LOLA slope map is shown in Figure 7.

The
closest slope measurement (at 172 m from the impact location) had a
slope of -0.14$\,^{\circ}$ in the East-West axis and 1.83$\,^{\circ}$ in the North-South axis, where a positive slope (x, y)
dips (eastward, northward) -- thus the total slope was 1.83$\,^{\circ}$. The average of the 6 closest points (see Appendix D)
were -1.00$\,^{\circ}$ in the East-West axis and 1.04$\,^{\circ}$ in
the North-South, with the standard deviation being 2.20$\,^{\circ}$
and 0.80$\,^{\circ}$ respectively. Thus at the Centaur impact site the slope is most likely dipped towards the North and West, (mainly North), with slopes $-0.14 \pm 2.2$ and $1.83\pm 0.8\,^{\circ}$. This does not account for any gradients with length scales $<400$m.

Combining the two measurements, and assuming a flat terrain at a $<400$m scale, we have that the impact angle with respect to the local slope $0.33 \pm 0.8\,^{\circ}$
in latitude and $3.66 \pm 2.2\,^{\circ}$ in
longitude.

In both directions the slope tends to cancel the local impact
velocity. The overall surface impact angle is thus $3.67 \pm 2.3\,^{\circ}$
in a mainly East direction. Thus one possible explanation for why the images of
the Centaur impact crater seen in the NIR2 camera show a greater amount of ejecta in the Easterly direction that the impact angle resulted in an asymmetrical plume with more ejecta in that direction. On the other hand, this does not seem to be consistent with models and ground based experiments which show little variation of ejecta symmetry with impact angle if one used high impact angles (close to normal). In general, it is perhaps worth noting that this impact angle is somewhat steeper than the pre-encounter estimations used in impact models and lab models (of order 70 degrees), see e.g. \cite{shuvalov2008, korycansky2009}, 

\subsection{Impact Angle Error Analysis}
The errors are dominated by the local slope errors. Given that the
closest measured slope is some 172 m from the Centaur impact
location, and given that the slope varies by of order 2$\,^{\circ}$
based on translations of that scale between different LOLA slope
measurements, in the region close to the impact point, the error margins on this
basis are approximately 2$\,^{\circ}$. The standard deviation of the 6 local slopes gives
$0.8\,^{\circ}$ in the N-S direction and $2.2\,^{\circ}$ in E-W
direction. Meanwhile the $1\sigma$ errors  in the angle deriving from the trajectory errors are negligible at $<0.01\,^{\circ}$. 

% * \section{Mission Requirements $\&$ Lessons Learned}
\section{Mission Requirements $\&$ Lessons Learned}
\label{sec:requirements}

Table \ref{requirements-table} lists six LCROSS mission requirements
related to the Centaur and the SSC impact locations, the impact
angles and observing geometry, and provides a summary of how well
they were met based on analysis herein.

\begin{table}[b]
\caption{ Six key related mission requirements of LCROSS how they were met}
\begin{tabular}{ l p{5.5cm} p{4cm}}
\hline
Req. No.&Requirement Text&Result\\
\hline
LPRP4.4	&The LCROSS impactor shall impact the target location at an
angle greater than 60 degrees to the plane of the lunar surface.
&Impact angle $85.92 \pm 0.01\,^{\circ}$ w.r.t. local slope.\\
PRJ4.2.3 &The LCROSS shall be able to target a 10 km radius impact
area (3-sigma). &Distance to target: 146 m, max. mean plus $3\sigma$ range $<500$m\\
PRJ6.4.8 &LCROSS shall be capable of observing impact target location and the sunlit ejecta cloud with at least 0.1 Hz frequency from EDUS impact minus 5 minutes until EDUS impact minus 3 minutes&Sample rates varied by instrument at time, but all were $<0.1$ Hz over this period and were observing the target location and ejecta cloud.\\
PRJ6.4.14 &The FS shall maintain payload pointing to the Centaur
impact site at the time of Centaur impact with an accuracy of +/-
0.5 degrees &Pointing error $< 0.1\,^{\circ}$ at impact\\
PRJ6.4.15 &The FS shall maintain payload pointing to the Centaur
impact site at Centaur impact plus 60 seconds with an accuracy of
+/- 0.64 degrees.  &Payload pointing to impact location
$<0.1\,^{\circ}$ at impact plus 60s\\
PRJ6.4.16 &The FS shall point the nadir viewing NIR Spectrometer to
the Centaur impact site at a time of 60 seconds prior to S-SC impact
to an accuracy of +/- 5 degrees. & Payload pointing to impact
location $<0.2\,^{\circ}$ at SSC impact -60 s\\

\hline
\end{tabular}
\label{requirements-table}  %table10
\end{table}

Regarding PRJ 4.2.3, the project's informal targeting goal was 1.75
km, which was also met with a wide margin.  The approximately 100m-level targeting accuracy of the mission may be of interest not only for future lunar purposes but also for Near-Earth Object (NEO) missions, some of which are of order this size. There is no obvious reason why a repeat mission with the same capability avionics and tracking systems would not be able to achieve a comparable targeting accuracy. 

With respect to PRJ4.1.7 and PRJ6.4.14-16, the SSC kept the
Centaur impact location continuously within the $0.1\,^{\circ}$ of
the NSP1 field-of-view (FOV) from 5 minutes prior to Centaur impact
until 3 minutes after. Note that the commands controlling
pointing were generated from an impact site prediction made several
hours before the impact and the SSC stayed within $0.1\,^{\circ}$ of
that prediction.

The analysis herein might suggest potential ways one may have improved the mission design: 
\begin{enumerate}
\item{Centaur tracking improvements: one could have added some
  device such as a corner cube to enable laser ranging, or a
  transponder, to have enabled more
  precise tracking of the Centaur post separation from the SSC.}
\item{Camera pointing and distortion: several analyses could have
  benefitted from better ground based calibration using reference
  targets.}
\item{Reaction wheels: one of the main sources of error on the
  spacecraft trajectory was that due to errors in the use of
  attitude thrusters, which could have been nulled through use of
  reaction wheels for attitude control.}
\end{enumerate}

Naturally, these steps would most likely have had a cost and schedule implication which may have been prohibitive in this cost- and schedule-capped mission. Reaction wheels would have reduced
dispersions and also some operational risks, but would have forced
re-allocation of resources used elsewhere, increasing risks there, etc. 

It can equally be stated, that each of the mission objectives, and in
particular the central scientific questions, were answered with the mission as was.
In fact one can ask whether LCROSS was
over-designed in some areas and speculate on changes to the
mission that may have reduced complexity and cost:

\begin{enumerate}
\item{Avionics: potentially a simpler/less capable avionics system could
  have been employed since the targetting was met with such a high margin}
\item{Tracking accuracy: the tracking accuracy could have been less
  and still met the mission requirements, (e.g. DSN could have
  been used less). As specific examples, mission operations could have foregone the update to the pointing quaternions that was made $\sim5$ hrs before impact and the mission might have been able to use an earlier OD solution to generate the pre-separation load of quaternion and braking burn parameters in order to give more time to generate and do quality assurance on products.}
\item{Fuel for ACS:}
\begin{enumerate}
\item{less fuel could have been used for ACS while still meeting the
  pointing requirements by allowing larger attitude control margins of error}
\item{less fuel could be used by not doing force balancing while
  still meeting the spacecraft targeting requirements}
\end{enumerate}
\item{Fuel for TCMs: perhaps fewer Trajectory Correction Maneuvers
  (TCMs) could have been made}
\end{enumerate}

% * \section{Conclusion}
\section{Conclusion}
\label{sec:conclusion}

We hope that this paper supports understanding the LCROSS science
data.  In particular that:

\begin{enumerate}
\item The impact imagery registration and locations could aid follow-up observations from the
      ground and from lunar orbit, particularly from LRO e.g. for example the LRO-LOLA instrument to attempt more careful study of the topography surrounding the impact locations to try to identify the crater (and thus independently measure features);
\item The impact location, derived from several independent methods, may help to constrain or improve the map tie errors in the lunar coordinate reference systems. 
\item Full registration of all descent images from the automated bundle adjustment process used herein could provide additional,
      useful data to users of the LCROSS data and will be provided to the PDS;
\item The impact angle of the Centaur with respect to the terrain
      may aid the development of impact cratering and ejecta process models; and
\item The pointing accuracy and motion of the LCROSS spectrometer
      boresight may help users of the data to better understand the spectra.
\end{enumerate}

LCROSS mission requirements relating to targeting, instrument fields
of view, pointing accuracy and impact angle were all met.  In fact 5 of the 6 did so by better than a factor of 5. 

\section{Acknowledgements}
We would like to acknowledge the efforts and support of the following teams/entities without whom this analysis would not have been possible: the Mission Operations Support team at NASA-Ames Research Center (ARC), the Mission Design Team at ARC, the Maneuver Design Team at NASA-Goddard Spaceflight Center (GSFC), the Orbit Determination Team at the Jet Propulsion Laboratory, the Subsystem analysts and engineers at Northrup Grumman Space Technology and Northrup Grumman Technical Services and the operators of the DSN facilities at Goldstone, Madrid and Canberra.

%\begin{acknowledgements}
%If you'd like to thank anyone, place your comments here
%and remove the percent signs.
%\end{acknowledgements}

% BibTeX users please use one of
\bibliographystyle{aps-nameyear}      % basic style, author-year citations
\bibliography{example}   % name your BibTeX data base
\nocite{*}

% Non-BibTeX users please use

\newpage
\appendix
%\appendixpage

% * \section{Appendix A: Source Data and Analysis Process}
\section{Appendix A: Source Data and Analysis Process}
The data sources used for all analyses herein are described in overview in Table \ref{location-source-data-table}. The process flow, showing input data used for each impact location
estimation process shown schematically in Figures 8 and 9 for the SSC and Centaur respectively.

\begin{table}[h]
\caption{Source data for location measurements. Note that the LOLA data was produced in October 2011 based on $> 20$ months of integrated data and that the resolution of the DSN is 1m
  radial to the antenna (in the Earth direction) and 75m in orthogonal directions}
\begin{tabular}{ p {1 cm} p {1 cm} p {2.8 cm} p {1.3 cm} p {1 cm} p {2 cm}}
\hline
Type&	Source&Description&Area (km)& Resolution (m)&Time Stamp\\
%&&&(km)&(m)& \\
\hline
Terrain&LRO&LOLA terrain maps &$10\times 10$&$\sim 40$&Jan. 2010\\
NIR &LCROSS&NIR image&$4\times 5$&$\sim 10$&11:35:31.404\\
MIR&LCROSS&MIR image&$2\times 3$&$\sim 20$&11:35:31.401\\
Radar&Goldstone&Radar images before $\&$ after&$5\times 7$&$\sim 5$&May $\&$ Nov. 2009\\
Ranging/ Doppler&DSN&SSC ranging data taken up until 1.5hrs before impact&n/a&$1\times 75$  &$\sim1$/hour until 10:17:00.000\\
Attitude&LCROSS&SSC attitude sensors data via telemetry &n/a&$<0.1\,^{\circ}$&$\sim 1$Hz through descent\\
\hline
\end{tabular}
\label{location-source-data-table}
\end{table}

\begin{figure}[h]
\label{flowssc}
\begin{center}
\includegraphics[width=0.65\textwidth,angle=0]{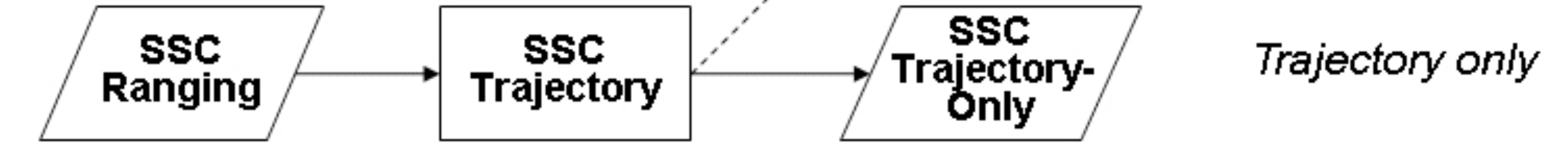}
\end{center}
\caption{Flow diagram of SSC impact determination process. On the
  LHS are the input data. On the RHS are the impact coordinate
  products. In the middle are
  the intermediate processes}
\end{figure}

\begin{figure}[h]
\label{flowcentaur}
\begin{center}
\includegraphics*[width=0.65\textwidth,angle=0]{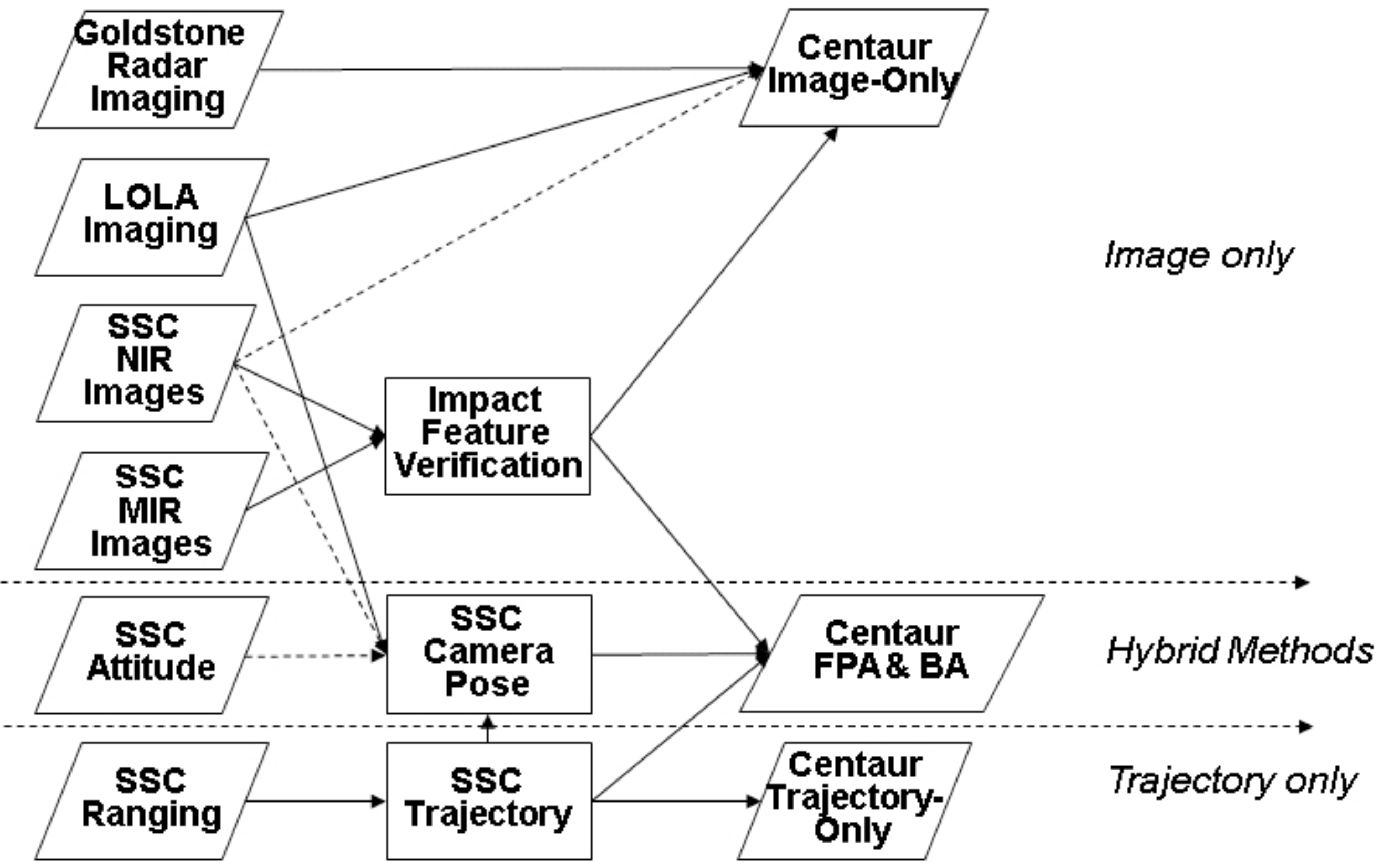}
\end{center}
\caption{Flow diagram of Centaur impact determination process. On
    the LHS are the input data. On the RHS are the impact coordinate
    products -- the four semi-independent methodologies (the image only and trajectory only are
    completely independent, and the manual and automated hybrid methods are semi
    independent of the other two). In the middle are the intermediate
    processes.}
\end{figure}

% * \section{Appendix B: The Centaur Impact Location}
\section{Appendix B: Further Details on The Centaur Impact Location}

In this section, each of the four Centaur impact location determination methods, as summarised in Table 2, are detailed. 

\subsection{Centaur Trajectory Only}

LCROSS tracking and orbit determination was performed by JPL using
transmitter ranging and Doppler data from the Deep Space Network (DSN).  With no onboard transmitter, the Centaur was untracked from the
time it separated from the SSC until impact 9.7
hours later.  Any trajectory-based estimate of the impact location
must address the trajectory perturbation introduced by the
separation event.  Two
methods of estimating the separation perturbation were used yielding two
estimates of the post-separation Centaur state vector.  These state
vectors were then propagated forward until they intersected the
surface.

% Will, I pulled the text from the appendix here.  I eliminated
% trajectory method 1 because you say it isn't applicable (and 

The first method involves using the results of pre-separation orbit
determinations to calculate the state vector of the combined
SSC/Centaur at the time of separation, then perturbing that vector
by the impulse applied (to each object) by the separation springs, as known from ground calibration, to derive the Centaur's post separation state vector.  This estimate of the
separation impulse is purely analytical.

The last state for the SSC prior to separation was calculated at 09
Oct 2009 00:20:00 UTC. The Cartesian coordinates in J2000 with the Earth
as the central body were:

\begin{itemize}
\item[] Position [km]: 87774.263494 340013.79800, 122539.97427
\item[] Velocity [km/s]: -0.84355753755, -0.09197672465, 0.85811369196
\end{itemize}

The separation event $\Delta V$ target was $ 0.721$ m/s with a $3\sigma$ uncertainties of 9 mm/s, per axis. The target attitude quaternion (in J2000) of the spacecraft at separation was: 0.36790183411, 0.73856335829, -0.40935830939, 0.38935611048. The spacecraft attitude quaternion as measured by the on-board attitude determination system at the time closest to separation (01:49:58.67) was -0.368135433852522, -0.738446417838845, 0.40900543141651, -0.389727786275225, suggesting an error of $<0.066 \,^{\circ}$ from the command (target) quaternion. The resulting Centaur post-separation state vector at 09 Oct 2009 01:50:00 is:

 % [Lateral velocity error?]

\begin{itemize}
\item[] Position [km]: 83216.293080, 339463.90550, 127193.59414
\item[] Velocity [km/s]: -0.84438148975, -0.11192890266, 0.86605975234
\end{itemize}

The second method involves using tracking data from before and
after separation plus Newton's Third Law.  Here the trajectory of
the combined SSC-Centaur vehicle prior to separation, together with
the trajectory of SSC post-separation (but prior to braking burn)
are taken, along with the SSC and Centaur relative masses at
separation. Together these enable one to deduce the post-separation
Centaur trajectory based on its equal and opposite reaction at the
separation event. The first state for the SSC post-separation was calculated for the
time 09 Oct 2009 02:28:59. The Cartesian coordinates in J2000 with
the Earth as the central body were:

\begin{itemize}
\item[] Position [km]: 81240.391374, 339191.93319, 129222.82223
\item[] Velocity [km/s]: -0.84480650128, -0.12083383916, 0.86981590338
\end{itemize}

The SSC mass at separation was 617.18 kg and the Centaur mass at separation 2271.61 kg. This, together with the SSC-Centaur pre-separation state already given above for the first method, were propagated to the separation time of 09 Oct
2009 01:50:00 -- in the case of the SSC post-separation state the propagation being
backwards in time. The resulting Centaur post separation state
vector, assuming an instantaneous maneuvre, and at the same time is:

\begin{itemize}
\item[] Position [km]: 83216.275714, 339463.90071, 127193.61830
\item[] Velocity [km/s]: -0.84444756135, -0.1118950345, 0.86592455048
\end{itemize}

The two post-separation Centaur state vectors were propagated forward until they intersected
the lunar surface using STK-Astrogator (9.0).  This propagator included a
70x70 LP-based lunar gravity model, the Sun and Earth as point
masses, as well as solar radiation pressure.  Based on LOLA data,
the surface altitude of -3.82633 km from a mean lunar radius of
1737.4 km. The JPL-based Orbit Determination team replicated the second method using an alternative propagation program to produce a
third, independent trajectory-based impact location estimate. The predicted impact locations from these methods and their average are below.

\begin{table}[h]
\caption{Summary results of the trajectory-only impact position estimates. Altitude is with respect to a sphere of radius 1737.4 km. Latitude and longitude are given in Mean Earth, selenocentric coordinates.}
\begin{tabular}{ l c c c}
\hline
Source&Latitude ($\,^{\circ}$)&Longitude ($\,^{\circ}$)&Altitude (km)\\
\hline
Trajectory 1&-84.6749921&-48.7001&-3.82817\\
Trajectory 2&-84.675&-48.703&-3.82633\\
Centaur JPL&-84.6749842&-48.6972&-3.83\\
\hline
Average&-84.6750&-48.7003&-3.82862\\
\hline
Average + outgassing&-84.6827&-48.6688&-3.82862\\
\hline
\end{tabular}
\label{trajectory-raw-table}
\end{table}

These estimates do not include a potentially significant,
intermittent perturbation: outgassing from ice remaining on the dark
side of the Centaur.  Deposited by the moist Florida air before
launch, trajectory perturbations from ice outgassing were observed
during a maneuver in August, two months after launch and one month
before the impact.  In the nominal LCROSS attitude, this ice is
shadowed by the Centaur and inert.  But after separation, the
Centaur tumbled, exposing the ice to the sun and causing
offgassing and consequently a small but unmeasured perturbation.

Dispersions due to ice outgassing will not be perfectly circular.
Since the outgassing occurs only when the ice is lit, it tends to
push the trajectory anti-sunward.  This effect is more pronounced
when the Centaur rotation period is less than the time before
impact, as it was.  Therefore, we expect outgassing to cause the
Centaur to land anti-sunward of the trajectory-based prediction
point.

The outgassing acceleration observed in August was 5.7 mm/s per
hour, applied to the mated SSC/Centaur ($\sim 3150$ kg).
Acceleration of the Centaur alone (2366 kg) would thus be approximately 7.6
mm/s per hour.  If these perturbations are assumed to be applicable
for half of the post-separation time (9.7 hours), then the total
deflection is 640 meters.  After applying cosine losses to
accelerations during Centaur rotation, the impact deflection can be
bounded to be $> 410$ m.  The lower limit is 100m.  Thus the
average would be 250m with an approximately 50m $1\sigma$ error to
account for these $3 \sigma$ bounds. These are the dominant contributions to the errors given in the main text (the propagation and initial condition errors being smaller). The results from these methods are summarised in Table 7.

% *** \subsubsection{Centaur Trajectory Only Error Analysis}
\subsubsection{Centaur Trajectory Only Error Analysis}

The standard deviation of the trajectory methods is 89m in latitude
and 107m in longitude.  However, the standard deviation of this
clustering does not include any systematic offset of this
methodology.  To understand the true error, one needs a bottom up
approach to error estimation.

The determination and prediction of the LCROSS trajectory was
performed by the Orbit Determination (OD) team at JPL.  Based on a
solution containing tracking data up to the separation event, JPL's
assessment of the $1\sigma$ knowledge of the Centaur Impact is a
nearly circular ellipse with radii of 94 meters.  This is primarily
attributable to three sources of uncertainty: (1) the state
knowledge at the time of the last tracking data observation of the
Centaur trajectory (i.e. just before separation).  Specifically the
cross-track knowledge at 36 meters and 2.3 mm/s, $1\sigma$,
equivalent to a deflection of 88m; (2) the effects of the separation
$\Delta V$ uncertainties on the relatively massive Centaur.
Specifically these were 3 mm/s, per axis, $1\sigma$, equivalent to a
deflection of 105m; and (3) the errors of potential outgassing as
residual water ice is burned off whenever the Centaur dark side
tumbles into the sunlight.  The uncertainties on this amount to 50m
$1\sigma$. Note that these need to be applied around a position that is offset by the mean expected offgassing. Adding these errors together we find a total error of $141\times 141$m.

% ** \subsection{Centaur Imagery Only}
\subsection{Centaur Imagery Only}

The method for locating the impact based on imagery only uses the
co-registration of impact crater images described in Section \ref{sec:registration}. Since LOLA
data is registered (in the Mean Earth frame), the process of 
registering the NIR2, MIR1 and GSSR images (which show the Centaur crater) to LOLA allows us to estimate
the Centaur impact location in absolute terms. The crater feature is most clearly visible in the NIR2 image. The location of the centre of this feature is -84.6774$\,^{\circ}$, -48.691$\,^{\circ}$ in Mean Earth, selenocentric coordinates.

Note that with the data used herein, it has not been possible to identify the Centaur crater
in the LOLA data but further data may allow the crater to be resolved. Note also that the feature in the NIR2 imagery which is over 150 m in diameter, is dominated by the bright ring of the freshly ejected material around the Crater, which has different scattering properties, but is likely to make little difference topologically, so the LOLA feature would be expected to be considerably smaller, given the crater size of 25-30m \citep{schultz2010}.

% This is due to three potential reasons. Firstly, the LOLA lateral resolution is approximately 40m for the data products used in this analysis whereas the crater produced by the impact is approximately 20 m in diameter. Secondly, the data is a mix of pre and post impact LIDAR measurements. Thirdly, the actual crater is quite shallow (approximately 2 meters), and that is close to the vertical resolution of these data. The feature in the NIR imagery is dominated by the bright ring of the freshly ejected material around the Crater which has different scattering properties, but will make little difference topologically.

% NOTE: What is the LOLA resolution?

% *** \subsubsection{Centaur Image-Only Error Analysis}
%\subsection{Centaur Image Only Error Analysis}

\subsubsection{Image Only Method Error Analysis}

%\begin{table}[ht]
%\caption{Image only raw impact location measurements for Centaur. Latitude and longitude are given in Mean Earth, selenocentric coordinates.}
%\begin{tabular}{ l c c}
%\hline
%&	Latitude ($\,^{\circ}$)&	Longitude ($\,^{\circ}$)\\
%\hline
%NIR-based&-84.6774&-48.7399\\
%\hline
%Average&-84.6731&-48.7304\\
%Standard dev. &0.000377&0.00348\\
%\hline
%\end{tabular}
%\label{image-only-raw-impact-table}
%\end{table}

The error for this approach is given by three factors. First the
error in the image registration.  This was estimated by repeating the
whole registration process and measuring the difference in impact
location. The maximum range of trials was found to be $\sim 100$m in latitude and longitude.  Second the error associated
with measuring the impact site in the NIR2 image.  This was estimated
by taking repeated measurements of a given registered image set,
outlining the errors.  Further, depending on where the impact point
is within the features, the errors build.  The NIR2 feature is
100x120m.  The overall error associated with where the
impact point in these features is approximately 70x70m.  Finally,
there is the accuracy with which the LOLA data has been registered to the lunar coordinate frame (the map tie error), which is approximately 40m.  These errors
are independent so we combine them using a quadratic mean to give
130m. This is approximately equal in both axes.

% ** \subsection{Hotspot Forward Projection}
\subsection{Hotspot Forward Projection (Manual)}

Two more approaches were developed which used a combination of
imagery and tracking data to determine the impact location.  The
first involves using Satellite ToolKit (STK) to project the vector
through the hotspot in the MIR1 images from the SSC (knowing it's attitude and position) to the surface.
The second method is a semi-automated image registration process
that was applied to all of the images.  This process was also
used to generate image registration data to be contained in supplemental data to the LCROSS
delivery to the Planetary Data System (PDS).

The Centaur impact location is clearly visible in MIR1 (thermal)
images taken in the last 30 seconds of the mission.  For each image,
the first technique projected a ray from the SSC centre's position when the
image was captured, through the hotspot in the image plane of the camera to the
surface.  This produced a cluster of points (corresponding to different MIR1 image times) on the surface whose
locations were averaged to estimate the Centaur impact location. The information needed to implement this method is: (1) the SSC
trajectory; (2) the SSC attitude history; (3) the MIR1 camera attitude
relative to the SSC body coordinate system; (4) the pixel locations
of the hotspots in the MIR1 images; and (5) the altitude of the
lunar surface terrain near the impact location.

This approach was implemented using STK and the LCROSS DataBrowser.
STK had already been used to model the SSC trajectory and attitude.
Within STK, the Moon was modelled as a sphere of radius 1737.4 (lunar
mean radius) with a texture generated from LOLA terrain data.  Also,
for the purposes of this analysis, the area surrounding
the Centaur impact was modelled as being flat at an altitude of -3.82633 km from the
lunar mean radius.  Therefore, STK took care of items 1, 2 and 5.

The LCROSS DataBrowser is custom software that we use to explore
payload imagery and spectra.  Here, it's simply a program for
selecting between MIR1 images along a timeline and displaying them
in a window.  The DataBrowser was modified so that the window
displaying the images could be made partially transparent so that
STK windows could be seen underneath.

Because no images show the Centaur crater together with enough
context to match against the LOLA-derived image, the camera pointing was
generated from an image taken $\sim 20$ minutes earlier and then
reused for later images that show the Centaur crater hotspot.  The
pointing was generated by selecting an image in the DataBrowser,
transferring the time to STK to generate an image from the S-S/C
position and attitude at that time, then adjusting the both the STK
image (varying rotation around the boresight) and the DataBrowser
image (adjusting X and Y scale and position) so the images matched.
%Figure \ref{pixvecscreen}.1 shows the result.

%\begin{figure}[ht]
%\label{pixvecscreen}
%\begin{center}
%%\includegraphics*[width=12cm,angle=0]{pixvecscreen}
%\includegraphics*[width=12cm,angle=0]{centaur-impact-alignment1}
%\end{center}
%\caption{An STK-based image registration process: LCROSS images are displayed on top of STK by a second program.  The overlay is
%  partially transparent so STK's coordinate system and an
%  LRO-derived image be seen through it allowing the images to be
%  aligned manually.  Once aligned with the LCROSS images, STK
%  natively provides the ability to find the location of an image
%  feature.}
%\end{figure}

This process generates the pointing for the camera implicitly, i.e., it
doesn't generate the pointing as a quaternion or transformation matrix,
but it does align the two programs so that images at other times
will remain aligned.  Once this has been done, later images showing
the hotspot can be superimposed over the STK display.  Within STK, a
target object was created at an altitude of -3.82633 km and moved in
latitude and longitude to lie under the hotspot.  This process was
repeated for each of five MIR1 images resulting in 5 targets whose
positions were averaged.  The entire process was then repeated from
the beginning. The results from these methods are summarised in Table 8.

\begin{table}[t]
\caption{Impact location estimates resulting from one iteration of
  the manual image registration method. Latitude and longitude are given in Mean Earth, selenocentric coordinates.}
\begin{tabular}{ l c c}
\hline
MIR1 image &	Latitude ($\,^{\circ}$)&	Longitude ($\,^{\circ}$)\\
\hline
Measurement 1&-84.6820039&-48.74303581\\
Measurement 2&-84.68174335&-48.74546489\\
Measurement 3&-84.68177716&-48.75053408\\
Measurement 4&-84.68163383&-48.75641848\\
Measurement 5&-84.68171144&-48.75977728\\
\hline
Average&$-84.68177394 \pm0.000139$&$-48.75104611\pm0.00708$\\
\hline
\end{tabular}
\label{manual-raw-table}
\end{table}

% *** \subsubsection{Hotspot Forward Projection Error Analysis}
\subsubsection{Hotspot Forward Projection Error Analysis}

Errors in the Hotspot Forward Projection Method arise from (1) the
SSC trajectory, (2) the SSC attitude, (3) MIR1 camera pointing, and (4)
alignment of surface targets with the hotspot in the MIR1 images.
Errors in camera pointing arise from (5) manual image alignment and (6)
registration of the LOLA terrain, to which an MIR1 image was
aligned, to STK's Mean Earth coordinate frame.  \#6 was trivial
compared to \#5 because the SSC was at a high altitude when the MIR1
image was taken.

We divide these into errors arising in the trajectory estimate and
all errors arising in the image alignment and attitude.  The
different image-related error sources are complex, so we estimate
the error by assuming a 2-D normal distribution for the resulting
impact crater locations.  From the 10 samples, the $1\sigma$ error
is 18 m.  The SSC trajectory error is 3 m x 75
m, $1\sigma$ in the lunar surface plane, where 3 m is in the Earth vector, and the other is
orthogonal.  These two error sources are independent, so we combine
them with their quadratic mean (RMS), handling each axis separately.
This yields an error ellipse for manual image registration of 21 m x
77 m.  Including the 40 m LOLA map-tie error, this goes to 45 m x 87
m, $1\sigma$, where the smaller error is along the Earth-axis and the larger
is orthogonal to that.

% ** \subsection{Automated Image Registration by Bundle Adjustment}
\subsection{Image Registration by Bundle Adjustment (Automated)}
\label{sub:BundleAdjustment}

The Bundle Adjustment (BA) approach was applied to solve for the relative pose and projective properties of the five LCROSS cameras with respect to the SSCÕs frame. With that solution, all LCROSS images where projected onto the LOLA topography where the latitude and longitude of individual pixels could be extracted. Implementation and algorithm were taken from \citep{Broxton}.

This algorithm reduces the projection error of 3D features into cameras by comparing the observed location of features in the images compared to where the camera models predicts them. The location of the 3D features were solved simultaneously along with the the rotation of the cameras with respect to SSC and their individual focal lengths. Two types of features where used in the optimization, Tie Points and Ground Control Points. Tie points are features found in 2 or more images where only the their pixel location was recorded. Their 3D location was seeded with a a mid-point triangulation method. Ground Control Points (GCPs) are image measurements similar to Tie Points that also have been registered to LOLA topography by hand. GCPs allow for an additional cost function in BA where their solved 3D location must stay close to their observed location on LOLA.

Two time instances where found where the five cameras (VIS, NIR1, NIR2, MIR1 and MIR2) were all taking exposures within a couple of seconds. 9 GCPs were tracked in those 10 images simultaneously. An additional 58 Tie Points were tracked but with varying coverage among the 10 images \citep{moratto2010}. The results from these methods are summarised in Table 9.

% ** \subsection{Hybrid Methods}
%\subsection{Hybrid Methods}
% *** \subsubsection{Automated Image Registration (by Bundle Adjustment)}
%\subsubsection{Automated Image Registration (by Bundle Adjustment)}

%See Appendix D.

\begin{table}[h]
\caption{Summary of measurements of impact location by automated bundle adjustment. Latitude and longitude are given in Mean Earth, selenocentric coordinates.}
\begin{tabular}{ l c c}
\hline
MIR1 image &	Latitude ($\,^{\circ}$)&	Longitude ($\,^{\circ}$)\\
\hline
1&	-84.676776&	-48.7064\\
2&	-84.676755&	-48.7036\\
3&	-84.67627&	-48.7098\\
4&	-84.677526&	-48.7001\\
5&	-84.676863&	-48.7177\\
6&	-84.676997&	-48.7006\\
7&	-84.677283&	-48.7031\\
8&	-84.676841&	-48.7149\\
9&	-84.680077&	-48.7204\\
10&	-84.679855&	-48.7263\\
11&	-84.679866&	-48.7396\\
12&	-84.679064&	-48.742\\
13&	-84.680799&	-48.7561\\
14&	-84.680985&	-48.7594\\
\hline
Average&	$-84.6783 \pm0.0017$&$	-48.721 \pm0.020$\\
$1\sigma$ error (m)&	52&	57\\
\hline
\end{tabular}
\label{bundle-raw-table}
\end{table}

% *** \subsubsection{Error Analysis}
\subsubsection{Error Analysis}

Errors and uncertainty where taken from derivatives of the last iteration of Bundle Adjustment. These translate to a ground position error of 52 x 57 m when projecting images near impact. In addition, LOLA 1$\sigma$ map tie errors are approximately 40m. The trajectory errors are 3x75m. Thus a total estimate might be 66x102m 1$\sigma$, where the smaller axis is in the Earth vector (projected onto the lunar surface) and the larger is orthogonal \citep{moratto2010}

\newpage
\section{Appendix C: Further Details on the SSC Impact Location}

% ** \subsection{SSC Trajectory Only}

In this section, the SSC impact location determination method based on the trajectory only is detailed. The SSC was tracked until quite close to impact, providing a fairly
accurate location.  The last state vector available for the SSC
prior to impact is for 09 Oct 2009 10:17:00 (approximately 79
minutes before impact).  The Cartesian coordinates in J2000 with the
Earth as the central body are:

\begin{itemize}
\item[] Position [km]: 57759.363734, 333782.02444, 155582.61612
\item[] Velocity [km/s]: 0.798778319, -0.327045732, 1.16015291
\end{itemize}

This state was propagated by STK-Astrogator (9.0) until the
trajectory met the lunar surface.  Astrogator used a 70x70 LP-based
lunar gravity model, the Sun and Earth as point masses, and included
solar radiation pressure and general relativity.  The altitude cut
off was taken to be -3.82633 km from a mean lunar radius of
1737.4km, based on LOLA data.  The impact position determined for
the spacecraft was: Latitude, Longitude, altitude: -84.719
$\,^{\circ}$, 49.610 $\,^{\circ}$, -3.80909 km.

The errors on the SSC trajectory as calculated by the JPL Orbital Determination team are 3 m x 75 m, $1\sigma$,
where the 3m is in the Earth vector (as projected onto the lunar
surface) and the 75 m is orthogonal to that. These are dominated by
the errors of the tracking process.

% ** \subsection{SSC Imagery Only}

% * \section{Appendix D: Bundle Adjustment}
%\section{Appendix D: Bundle Adjustment}

%[Full details from Zach's report?]

% * \section{Appendix E: Local Slope Measurements}
\newpage
\section{Appendix D: Local Slope Measurements}

Table 10 shows slope measurements made by LOLA that are local to the Centaur impact site. These are used in Section 7.
 
\begin{table}[h]
\caption{Local Slope measurements. (Note the impact location is -84.6775, -48.7216)}
\begin{tabular}{ p{1.5cm} p{2cm} p{2cm} p{1.5cm} p{1.5cm} p{2cm}}
\hline
Slope&Latitude ($\,^{\circ}$) &Longitude ($\,^{\circ}$)&E-W slope ($\,^{\circ}$)&N-S slope ($\,^{\circ}$)&Impact Distance (km)\\
\hline
Point 1	&-84.6566&	-48.75339&	-3.01871	&0.691863&	0.649182\\
Point 2&	-84.6615&	-48.75	&-0.01793	&-0.24683	&0.497813\\
Point 3&	-84.6676&	-48.76142&	-4.26773	&1.646793&	0.321277\\
Point 4	&-84.6726&	-48.75794	&-0.14788&	1.828963&	0.172431\\
Point 5	&-84.69&	-48.77573	&-0.2415	&1.588041&	0.38113\\
Point 6	&-84.6944&	-48.76999	&1.67379	&0.71209&	0.50838\\
\hline
Average	&-84.6738&	-48.76141&	-1.00333&	1.03682\\	
Standard Dev.	&0.015325&	0.0098531&	2.198553&	0.796945	\\
\hline
\end{tabular}
\label{local-slope-raw-table}  %table9
\end{table}

\end{document}